\documentclass[structabstract]{aa}  
\usepackage{graphicx}
\usepackage{txfonts}
\usepackage{natbib}
\begin{document}
   \title{On the possibility of detecting extrasolar planets' atmospheres with the Rossiter-McLaughlin-effect} 

%   \subtitle{Exoplanets' atmospheres and the RME}

   \author{S. Dreizler
          \inst{1}
          \and
          A. Reiners\inst{1}
          \and
          D. Homeier\inst{1}
          \and
	  M. Noll\inst{1,2}
% \thanks{}
          }

   \institute{Institut f\"ur Astrophysik, Georg-August-Universit\"at
              G\"ottingen, Friedrich-Hund-Platz 1, D-37077 G\"ottingen, Germany\\
              \email{dreizler@astro.physik.uni-goettingen.de} \and
	      Gymnasium am B\"otschenberg, Am B\"otschenberg 11,
              D-38350 Helmstedt 
             }

   \date{Received ; accepted}

\abstract{The detection of extrasolar planets' atmospheres requires
  very demanding observations. For planets that can not be spatially
  separated from their host stars, i.e. the vast majority of planets,
  the transiting planets are the only ones allowing to probe their
  atmospheres. This is possible from transmission spectroscopy or from
  measurements taken during secondary eclipse. An alternative is the
  measurement of the Rossiter-McLaughlin-effect, which is sensitive to
  the size of the planetary radius. Since the radius is
  wavelength-dependent due to contributions of strong planetary
  absorption lines, this opens a path to probe planetary atmospheres
  also with ground-based high-resolution spectroscopy.}
{The major goal of our
  numerical simulations is to provide a reliable estimate of the
  amplitude of the wavelength-dependent Rossiter-McLaughlin-effect.}
%{Due to absorption within a planetary atmosphere, the radius of the
%  planet is wavelength-dependent with a variation of the order of
%  5\%. It is most prominent at wavelengths of strong absorption lines,
%  e.g. the NaD-lines. We can therefore use the wavelength dependency
%  of the Rossiter-McLaughlin-effect to analyse planetary atmospheres.}
{Our
  numerical simulations provide phase resolved synthetic spectra
  modeling the partly eclipsed stellar surface during the transit in
  detail. Using these spectra we can obtain Rossiter-McLaughlin-curves
  for different wavelength regions and for a wavelength-dependent
  planetary radius. Curves from regions with high and low contributions
  of absorption lines within the planetary atmosphere can be
  compared. From these differential effects observable quantities are
  derived.}
{We applied our simulations to HD\,209458. Our numerical simulations
  show that a detailed treatment of the limb darkening for the
  synthetic spectra is important for a precise analysis. Compared to
  a parameterized limb darkening law, systematic errors of
  6\,m\,sec$^{-1}$ occur. The wavelength dependency of the planetary
  atmospheres over the NaD-doublet produce a differential effect in
  the Rossiter-McLaughlin-curve of 1.5\,m\,sec$^{-1}$ for a star with
  a rotation velocity of $4.5$\,km\,sec$^{-1}$ which increases
  to 4\,m\,sec$^{-1}$ for twice the rotation velocity.} 
{The Rossiter-McLaughlin-effect as tool to probe planetary
  atmospheres requires phase resolved, high signal-to-nose,
  high-resolution spectra taken with a stabilized spectrograph in
  order to obtain reliable results for slowly rotating ($<10$\,m\,sec$^{-1}$)
  planet host stars. Stars with spectral type earlier than about F5
  are a bit less demanding since the typically higher rotation
  velocity increase the amplitude of the effect to about
  15\,m\,sec$^{-1}$ for a star with $v\sin i=25$\,km\,sec$^{-1}$.}
 

   \maketitle
%
%________________________________________________________________
\section{Introduction}

Within the recent two decades, the search for extrasolar planets has
been mainly driven by the success of the radial velocity (RV)
method. A large fraction of our knowledge about the physical
properties relies, however, on the analysis of transiting
planets. This is due to the fact that for transiting planets more
information can be obtained from various techniques: The radius and
the orbital inclination from the light curve analysis together with
the mass function from RV-measurements provide the mean planetary
density, an important constraint for the structure and evolution of
the planet. The atmosphere can be probed by means of transmission
spectroscopy during transits \citep{2002ApJ...568..377C,
2003Natur.422..143V, 2004ApJ...604L..69V, 2007ApJ...661L.191B,
2007Natur.448..169T, 2008A&A...483..933E, 2008A&A...481L..83L,
2008A&A...485..865L, 2008MNRAS.385..109P, 2008ApJ...673L..87R, 
2008ApJ...686..658S, 2008ApJ...686..667S, 2008A&A...487..357S,
2008Natur.452..329S}, the 
albedo and thermal emission can be determined from the secondary
eclipse with infrared photometry \citep{2005ApJ...626..523C,
2005Natur.434..740D, 2006ApJ...644..560D, 2007ApJ...667L.199D,
2007A&A...475.1125D, 2007ApJ...658L.115G, 2007Natur.447..691H,
2007Natur.445..892R,2008ApJ...674..482S}.

From eclipsing binary systems it is known since decades, that the
radial velocity curve shows a characteristic feature during eclipses
\citep{1910PAllO...1..123S, 1924ApJ....60...15R,
1924ApJ....60...22M}. This Rossiter-McLaughlin-effect (RME) can be
explained by the partial coverage of the eclipsed star. Out of
eclipse, the spectral lines will be symmetric since all surface
elements of the (rotating) star contribute equally, thus blue and red
shifted regions of the surface are balanced. This is changed during
the eclipse depending on the path of the occulting body with respect
to the eclipsed star, resulting in asymmetric lines due to unbalanced
Doppler-shifted contributions. Depending on the spectral resolution of
the spectrograph and the rotational velocity of the star this can be
either detected as pseudo RV variation (slow rotation or low
resolution) or directly as line distortion moving over the spectral
lines during the eclipse. The RME has been shown to be a powerful tool
for transiting planets as well. The method has been applied for the
measurement of the spin-orbit-alignment for eleven of the transiting
planets \citep{2000PASP..112.1421B, 2000A&A...359L..13Q,
  2005ApJ...631.1215W, 2006ApJ...653L..69W, 2007ApJ...665L.167W,
  2007PASJ...59..763N, 2007ApJ...667..549W, 2008A&A...482L..25B,
  2008ApJ...683L..59C, 2008A&A...488..763H, 2008ApJ...686..649J,
  2008arXiv0806.1478J, 2008A&A...481..529L, 2008PASJ...60L...1N}. With
the fast growing number of bright host stars of transiting planets,
either from ground-based wide field surveys or from space missions,
this measurement turns into a standard tool for investigating
properties of extrasolar planets.

The RME was also proposed as tool for probing planetary
atmospheres with ground-based observations
\citep{2004MNRAS.353L...1S}. Since transmission spectroscopy as
well as albedo measurements mainly rely on the high
precision obtainable only from space-based measurements (see
\cite{2008ApJ...673L..87R, 2008A&A...487..357S} as exception),
ground-based access is an attractive alternative. The RME therefore is 
a complementing rather than a competing method. The method detecting
planetary atmospheres with the
RME is based on the fact that the radius of an extrasolar planet
depends on wavelength. This is caused, on the one hand, by the
fact that the atmosphere of the planet shows variations in optical
depth with respect to the wavelength \citep{2007ApJ...661L.191B,
2008ApJ...678.1419F}. The planet will appear slightly larger in
optically thick spectral lines compared to wavelength regions where the
atmosphere is transparent. On the other hand, the amplitude of the RME
depends on the radius of the occulting body. A wavelength-resolved
measurement of that effect provides therefore an access to the
atmospheric properties of the transiting planet.

While the measurements by \cite{2004MNRAS.353L...1S} were at the limit
of the available data, it is the scope of this paper to provide
detailed modeling of the RME in order to provide a reliable analysis
tool for such demanding measurements. We therefore describe the
underlying model in Sect.\,2, and present test calculation for
HD\,209458 in Sect.\,3, the results are summarized in Sect.\,4.

\section{Modeling the Rossiter-McLaughlin-Effect}

Recently, there have been two papers on analytical expressions for the
RME, expressing the RV-shift in terms of orbital parameters and radius
of the planet, orbital orientation relative to the rotational axis,
stellar rotation velocity, radius as well as limb darkening
\citep{2005ApJ...622.1118O, 2006ApJ...650..408G} making use of earlier
analytical work \citep{1938PDAO....7..105P, 1953PASJ....5...88H,
1959cbs..book.....K}. While the analytical models are well suited as
fast and direct method, \cite{2005ApJ...631.1215W} argue that the the
first moment of the spectral lines, i.e. the shift in center of
gravity in wavelength space, as result of the analytical formula
might not be identical to the value measured from observations. In
that case, the optimal match between a template and the observed
spectrum is searched for.

The alternative approach therefore is a finite element model of the
star adding contributions from all surface elements to a
synthetic spectrum which can then be treated with the same analysis
tools as the observations, e.g. as used by \cite{2000A&A...359L..13Q}
and \cite{2005ApJ...631.1215W}. While computationally more demanding,
this approach allows to deviate from simplifications necessary in the
analytical models allowing for a more precise modeling of the RME. In
the following paragraphs the modeling of the wavelength-dependent RME
will be described.

\subsection{The star}
The assumption of a spherical primary star is well suited in the case
of a planetary secondary since deviations from spherical symmetry of
the primary due to tidal interaction is negligible. Spherical symmetry also requires
a slow rotation of the primary. For late type stars 
this assumption is well justified. It is, however, challenged when the
rotation velocity becomes significantly larger than in the Sun. In
order to prepare for such cases, we make use of the program {\it
BRUCE}, originally written by \cite{1997MNRAS.284..839T} to model
stellar pulsations in rotating stars. A deformation of the stellar
surface is calculated as perturbation providing a variation of the
effective temperature and surface gravity over the surface. As input,
the stellar mass, the polar radius and effective temperature, the
equatorial velocity as well as a velocity amplitude due to pulsation,
and the inclination $i$ of the rotation axis (assumed to be constant)
have to be specified. In our application, we set the pulsational
velocity to zero (this could, however, be included in order to
investigate a contribution of stellar pulsation to RV noise). In an
output file, {\it BRUCE} provides effective temperature, surface
gravity and projected radial velocity as well as the cosine of the
angle between the normal direction of the surface element and the line
of sight of the observer ($\mu$) for the required number of surface
elements (typically 50\,000 in our test calculations).

\subsection{Synthetic spectrum of the host star}
In previous modeling authors used template spectra, e.g. the solar
spectrum, or mean intensities from model atmospheres as input
spectra. In both cases, a limb darkening law
(e.g. \cite{2000A&A...363.1081C, 2003A&A...401..657C,
2004A&A...428.1001C}) has to be applied. For a detailed
wavelength-dependent modeling of the RME the available broad-band limb
darkening coefficients are not sufficient, instead this would require
determination of limb darkening coefficients from model atmospheres
adapted to the spectral resolution of the observations. In order to
avoid errors due to the parameterization of the limb darkening law,
specific intensities for all surface elements of the star can be used
directly for a synthetic spectrum.

The input spectra were synthesised using specific intensities (energy
emitted in a given direction $\vec{n}$ per time, unit area, frequency
and solid angle interval) from a grid of \texttt{PHOENIX} models.  The
grid covers the main sequence in steps of 200\,K for the effective
temperature and with two surface gravities (log\,g=4 and log\,g=4.5).
The models have been calculated for spherically symmetric atmospheres
under the assumption of local thermodynamic equilibrium
\citep{1999JCoAM.109...41H,1999ApJ...525..871H}. The general setup is
described in \citet{2005ESASP.576..565B}, however we are adopting the
revised solar abundances of \citet{2005ASPC..336...25A} to calculate
the chemical composition (for solar metallicity in the present case,
but being available for other metallicities as well).  Specific
intensity spectra for each model computed at a resolution of
0.01\,\AA\ for different values of the angle $\theta$ between the
normal direction of the atmosphere and the direction of the line of
sight ($\mu=\cos \theta$).  19 intensities per wavelength point are
extracted from all core rays and a subset of the tangent rays to
sample the range $0<\mu<1$ evenly.  Thus while our model treats the
global geometry of the stellar surface exactly including rotational
flattening, it is locally approximated in spherical geometry, while
interpolating to the exact effective temperature and gravitational
acceleration for each atmospheric segment. Since the deformation and
the total curvature are very small for dwarf stars, these
approximations lead to negligible errors.  The spectrum is then
synthesized by linear interpolation within the available grid of
intensities on a logarithmical equidistant wavelength grid of the
required spectral resolution and shifting to the required Doppler
velocity.  For comparison, we can also calculate intensities from the
disk-averaged spectrum by applying a linear limb darkening law. The
undisturbed spectrum of the planet host star is now obtained and can
be used as template for cross-correlations to determine the RME and it
serves as reference spectrum of the next step.

%As input spectra, we use specific intensities (energy emitted in a
%given direction $\vec{n}$ per time, unit area, frequency and solid
%angle interval) from a grid of {\it PHOENIX} plane-parallel model
%atmospheres. The grid covers the main sequence in steps of 200\,K for
%the effective temperature and with two surface gravities (log\,g=4 and
%log\,g=4.5). Specific intensities for each model are available in a
%resolution of 0.01\,\AA\ and for 19 values of the angle $\theta$
%between the normal direction of the atmosphere and the direction of
%the line of sight ($\mu=\cos \theta$). The metallicity of the star is
%specified via the metallicity of the model atmospheres (solar in our
%case). The spectrum is then synthesized by linear interpolations
%within the available grid of intensities on a logarithmical
%equidistant wavelength grid of the required spectral resolution. For
%comparison, we can also use mean intensities (averaged over the
%stellar disk) in combination with a linear limb darkening law. The
%undisturbed spectrum of the planet host star is now obtained and can
%be used as template for cross-correlations to determine the RME and it
%serves as reference spectrum of the next step.

\subsection{The Planet}
The RME is dependent on the size of the planet as well as on its
orbital parameters. Since transiting planets are typically close in,
the assumption of a circular orbit is well justified, however,
eccentric orbits with eccentricities above 0.1 have been detected in
four transiting planets: GJ\,436b, ecc=0.15 \cite{2007ApJ...667L.199D},
XO-3b, ecc=0.26 \cite{2008ApJ...677..657J}, 
HD\,147506b, ecc=0.5163 \cite{2007ApJ...670..826B}, HD\,17156b,
ecc=0.6753 \cite{2007ApJ...669.1336F}. We therefore 
specify the orbital separation and period. An extension towards
eccentric orbits would be possible in case this is needed. The
orientation of the planetary orbit with respect to the line of sight
towards the observer also has to be specified since the Doppler shifts
due to stellar rotation breaks the symmetry in this case. The
  orientation of the orbital rotation axis relative to the line of
  sight needs additional two parameters, the inclination $i$ and the
  tilt of the orbit in the plane perpendicular to the line of sight $\lambda$.

%While in
%other applications of the RME, the inclination of the planetary orbit
%relative to the apparent stellar equator is used as parameter
%(typically assigned as $\lambda$), we describe the spatial orientation
%of the orbital rotation axis relative to the line of sight with two
%angles $\alpha$ (orbital rotation axis with respect to the equatorial
%plane of the star) and $\beta$ (tilt of the orbit in the plane
%perpendicular to the line of sight). The angles $i$ and $\alpha$ are
%nearly degenerate, which means that 
%$\lambda$ and $\beta$ are nearly identical.

Synthetic spectra can now be calculated. We specify an equidistant
phase sampling (starting shortly before the first contact and ending
shortly after the fourth contact), the radius of the planet as well as
the angles $i$ and $\lambda$. For each orbital phase, the elements
on the stellar surface which are eclipsed are identified and their
spectra are interpolated as in the previous step. Their contribution
is then subtracted from the reference spectrum resulting in the
synthetic spectrum for the given orbital phase. This procedure can be
repeated for different planetary parameters.

\subsection{Wavelength-dependent planetary radius}
\begin{figure}
\centering \includegraphics[bb=54 520 558
720,width=8.8cm]{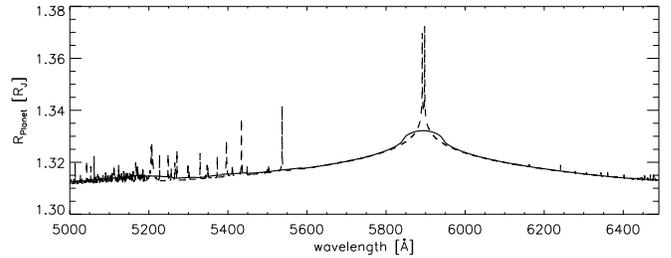}
\caption{Wavelength-dependent planetary radius of HD\,209458b
  \citep{2007ApJ...661L.191B} (dashed line) and smoothed over
  100\,\AA\ bins (full line).}
\label{FigBarman}
\end{figure}

\begin{figure}
\centering \includegraphics[width=8.8cm]{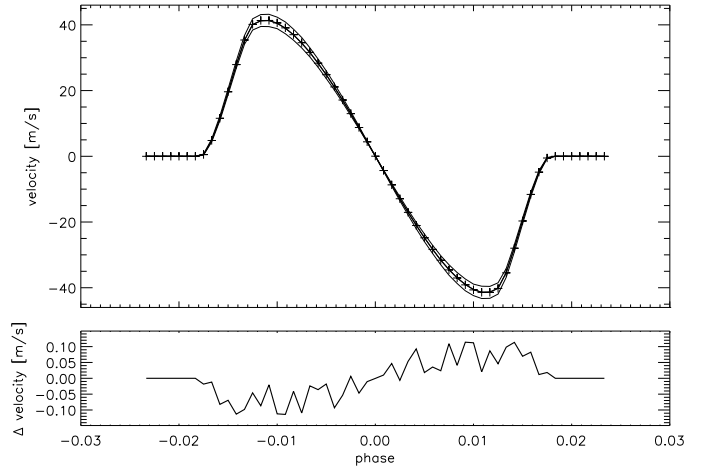}
\caption{Rossiter-McLaughlin-Effect for HD\,209458 (orbital motion
  subtracted): Top panel: Planetary radius R$_P$=1.37\,R$_{\rm Jup}$
  and R$_P$=1.31\,R$_{\rm Jup}$ (thin lines, full simulation),
  R$_P$=1.34\,R$_{\rm Jup}$ (thick line, full simulation) and
  R$_P$=1.34\,R$_{\rm Jup}$ (++++, interpolation procedure using
  R$_P$=1.37\,R$_{\rm Jup}$ and R$_P$=1.31\,R$_{\rm Jup}$). Lower
  panel: Difference between the full simulation and interpolation
  procedure for R$_P$=1.34\,R$_{\rm Jup}$. See text for details.}
\label{FigInterpol}
\end{figure}

From radiative transport calculations, \cite{2007ApJ...661L.191B}
deduce a wavelength dependency of the planetary radius as the result
of the varying optical depth of the planetary atmosphere (see
Fig.\ref{FigBarman} for an adopted version of his Fig.1). From his
results a variation in the order of 5\% can be expected, with a
prominent peak e.g. at the resonance lines of sodium.
Predictions of \cite{2008ApJ...678.1419F} indicate a variation between
2\% and 5\% depending on the class of irradiated atmosphere. In order
to take the wavelength-dependent radius into account, we calculate two
sets of synthetic spectra ($F_{\rm up}, F_{\rm low}$), one with the
upper limit and one with the lower limit of the radius ($r_{\rm up},
r_{\rm low}$). For each wavelength point, we then interpolate the two
synthetic spectra for the given wavelength point to the value
according to the prediction of \cite{2007ApJ...661L.191B},
i.e. $F_{\rm int,\lambda}=F_{\rm low}+(r_{\rm int,\lambda}-r_{\rm
low})\cdot (F_{\rm up}-F_{\rm low})/(r_{\rm up}-r_{\rm low})$. It
should be noted, that we are not restricted to these prediction but
can accommodate any given wavelength-dependent radius, e.g. based on
observations of transmission spectra \cite{2008ApJ...686..658S}. The
application 
of a linear interpolation is well justified as can be seen in
Fig.\,\ref{FigInterpol}. In the upper panel we display the RME for the
lower and upper limit of the predicted wavelength-dependent radius as
well as the RME for an intermediate value, one calculated directly as
described in the previous step the other one interpolated between the
two limiting cases. The lower panel shows the difference between the
two, which are of the order of cm\,sec$^{-1}$ and therefore certainly
below current detection limits.

\section{Results: Test case HD\,209458}
In the following, we present test calculations of the RME and transit
light curves for HD\,209458. The stellar and planetary parameters are
adopted from \cite{2004A&A...418..989N} and
\cite{2001ApJ...552..699B}, namely we use R$_{\rm \ast,
polar}=1.15$\,R$_\odot$, M$_\ast=1.05$\,M$_\odot$, v$\,\sin i=4.5$\,km\,sec$^{-1}$, T$_{\rm eff,polar}$=6000\,K, and an
inclination of the rotational axis of 90$^{\circ}$ as well as an
orbital separation of 0.047\,AU, an orbital inclination of
86.6$^{\circ}$ and no tilt of the orbit in the plane perpendicular to
the line of sight. %(in our notation $\alpha = 3.4^{\circ}, \beta=0^{\circ}$). 
The planetary radius is treated
wavelength-dependent (Fig.\ref{FigBarman}) as described in the
previous section using the predictions of \cite{2007ApJ...661L.191B}.

\subsection{The effect of limb-darkening}

\begin{figure}
\centering \includegraphics[width=8.8cm]{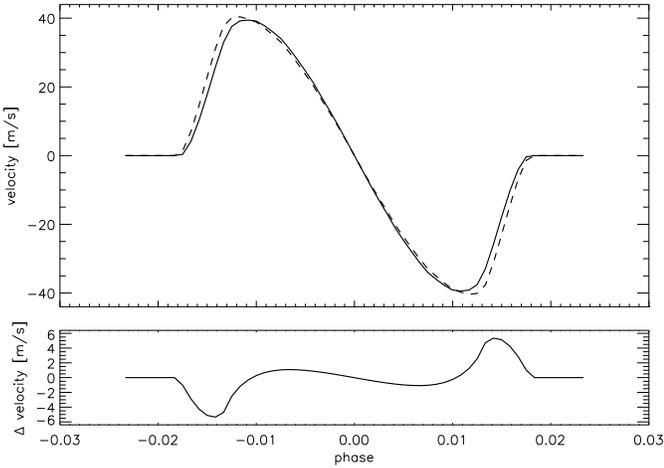}
\caption{The Effect of limb darkening: Top panel:
  Rossiter-McLaughlin-Effect (orbital motion subtracted) for the
  wavelength region 5000-5200\,\AA\ from synthetic spectra using
  specific intensities (full line) compared to the application of a
  linear limb darkening law (dashed line). Lower panel: Difference
  between the full simulation and the limb darkening
  approximation. See text for details.}
\label{FigDiff_ld}
\end{figure}

\begin{figure}
\centering \includegraphics[width=8.8cm]{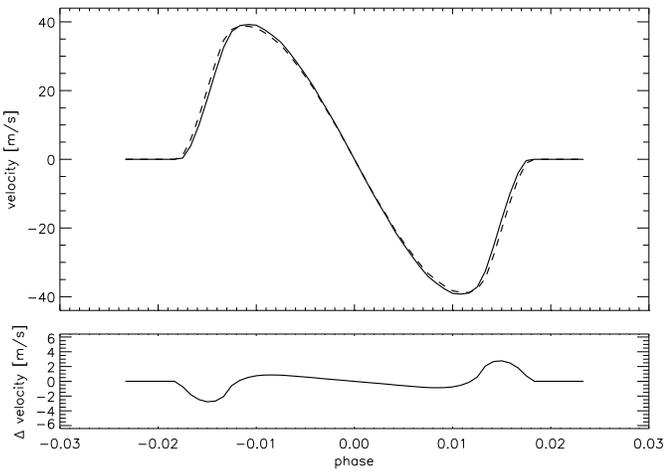}
\caption{As Fig.\,\ref{FigDiff_ld}, but for the application of a
  wavelength dependent linear limb darkening law shown in
  Fig.\,\ref{FigLimbdark}.} 
\label{FigDiff}
\end{figure}

\begin{figure}
\centering \includegraphics[bb=54 520 558
720,width=8.8cm]{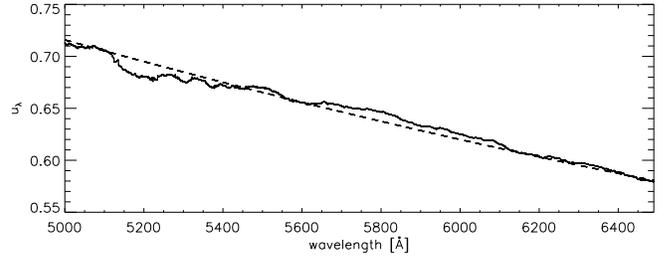}
\caption{Wavelength-dependent linear limb darkening coefficient
  determined from the applied {\it PHOENIX} models smoothed over
  100\,\AA\ bins (full line) and polynomial fit of second order
  (dashed line).}
\label{FigLimbdark}
\end{figure}

The form of the RME is not only dependent on the stellar and planetary
parameters, it is also sensitive to the treatment of limb
darkening. As described in Sect.\,2, our spectral synthesis relies on
the co-addition of specific intensities for each surface element,
interpolated to the actual value of the angle between the normal of
the element and the line of sight. In Fig.\,\ref{FigDiff_ld} and
\ref{FigDiff} we compare the averaged RME from a reference region
(5000-5200\,\AA) from our full simulation including specific
intensities to two versions of simulations using the flux and a linear
limb darkening law $ I_{\mu,\lambda} = 1+(\mu-1)u_\lambda $ for the
spectrum synthesis. In the first case we use a constant coefficient of
$u_\lambda=u=0.58$ used by \cite{2004MNRAS.353L...1S}, in the second
case we use a polynomial fit of second order to the
wavelength-dependent linear limb darkening coefficient
(Fig.\,\ref{FigLimbdark}). 

The deviation between the simplified limb darkening approach and the
detailed simulation using specific intensities is most evident when
the planet covers the stellar limb, i.e. before the second and after
the third contact. Here, the differences between the limb darkening
law and the actual $\mu$-dependence of the specific intensities are
largest. A limb darkening law therefore mocks a star with a different
effective radius. The differences can be reduced with a
wavelength-dependent limb darkening (Fig.\ref{FigDiff}) law but not
completely suppressed.

Systematic differences between the detailed treatment and a limb
darkening approximation are up to 6\,m\,sec$^{-1}$ in the first case
and 2\,m\,sec$^{-1}$ in the second case. A detailed treatment of the
RME might therefore help to reduce the error bars in determinations of
the spin-orbit alignments or misalignements of transiting
extrasolar planets 
\citep{2000PASP..112.1421B, 2000A&A...359L..13Q,
  2005ApJ...631.1215W, 2006ApJ...653L..69W, 2007ApJ...665L.167W,
  2007PASJ...59..763N, 2007ApJ...667..549W, 2008A&A...482L..25B,
  2008ApJ...683L..59C, 2008A&A...488..763H, 2008ApJ...686..649J,
  2008arXiv0806.1478J, 2008A&A...481..529L, 2008PASJ...60L...1N}. Due
to the wavelength 
dependency, a detailed treatment seems also necessary in order to
avoid confusion between effects of a wavelength-dependent planetary
radius and a wavelength-dependent limb darkening.

\subsection{The Wavelength-dependent Rossiter-McLaughlin-Effect}

\begin{figure}
\centering \includegraphics[width=8.8cm]{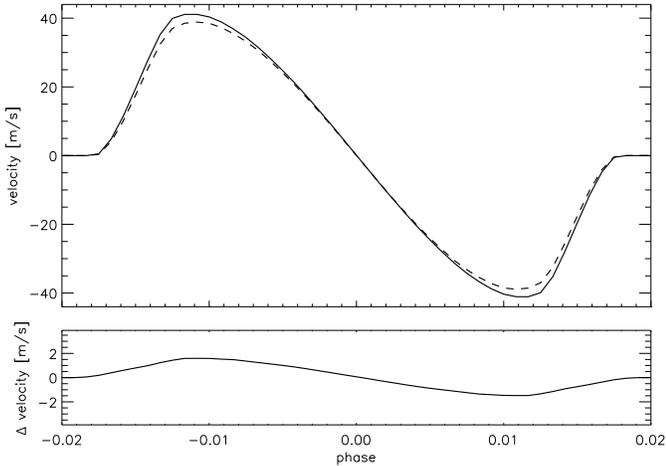}
\caption{Rossiter-McLaughlin-Effect (orbital motion subtracted) for
  the wavelength region 5840-5940\,\AA\, (100\,\AA\, bin centered
  around the NaD-doublet): Simulation with the wavelength-dependent
  radius for that wavelength interval (top panel, full line) versus a
  fixed radius R$_{\rm P}=1.31$\,R$_{\rm jup}$ (top panel, dashed
  line) and difference between the two (lower panel).}
\label{Figrmnld2}
\end{figure}

With our simulations and predictions for the wavelength dependency of
planetary radii \citep{2007ApJ...661L.191B, 2008ApJ...678.1419F} it is
now possible to simulate the wavelength dependency of the RME and to
estimate the observability. All test calculations were applied to
HD\,209458. It should, however, be noted that our simulations are
not restricted to that object. We assume a wavelength range from
5000-6500\,\AA, roughly covered by a spectrograph stabilized by an
iodine-cell and covering the NaD-doublet where the planetary radius is
increased by about 5\% due to absorption within the planetary
atmosphere. We used the phase-dependent synthetic spectra as described
in Sect.\,2, performed cross-correlations stepping through the
spectral region with a step size of 25\,\AA, and a width of 100\,\AA,
i.e.  averaging the RME over each 100\,\AA-bins. The correspondingly
averaged wavelength-dependent radius of HD\,209458b can be seen in
Fig.\,\ref{FigBarman}. As a reference, we use the RME averaged over
the region 5000-5200\,\AA, because the predicted radius is least
influenced by absorption in the planetary atmosphere in our spectral
region of interest. The differential effect is illustrated in
Fig.\,\ref{Figrmnld2}, where we show the RME of the reference region,
the 100\,\AA-bin centered around NaD, as well as the difference
between the two. The amplitude of the effect is about
1.5\,m\,sec$^{-1}$, about the quantity derived by
\cite{2004MNRAS.353L...1S}. It should be noted that the differential
effect is qualitatively different from that due to simplifications
using parameterized
limb darkening laws (Figs.\,\ref{FigDiff_ld} and \ref{FigDiff}). The
maximum effect occurs when the planet covers most of the blue- or 
red-shifted hemisphere, i.e. after the second and before the third
contact. The wavelength dependency of the planetary radius therefore
has a different signature in the RME than limb darkening. Using a
simplified limb darkening model as comparison for the observations
would in principle allow to discriminate between effects of the limb
darkening and the wavelength dependency, in practice the necessary
accuracy would be difficult to achieve. A detailed model is therefore
necessary to obtain reliable results.

\begin{figure}
\centering \includegraphics[width=8.8cm]{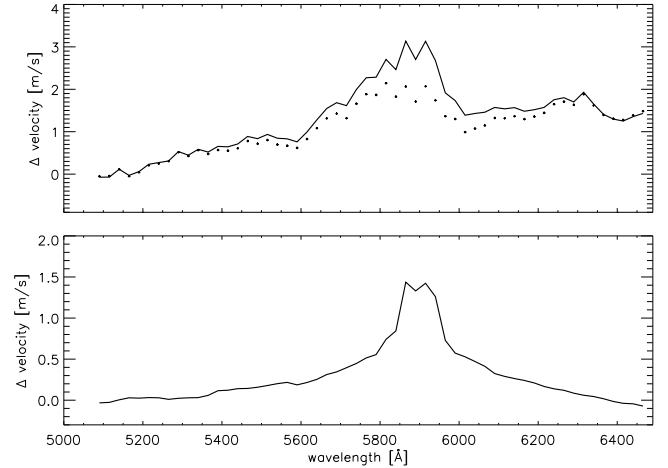}
\caption{Wavelength-dependent Rossiter-McLaughlin-Effect at the phase
  of maximum amplitude (-0.012): Top panel: Difference between the
  wavelength-dependent RME and the RME from the reference region
  5000-5200\,\AA\. Full line: Simulation with the wavelength-dependent
  radius, dotted line: Simulations with a a fixed radius R$_{\rm
  P}=1.31$\,R$_{\rm jup}$. Lowe panel: The difference between the two
  simulations from the top panel.}
\label{Figrmnld}
\end{figure}

In Fig.\,\ref{Figrmnld} we show the wavelength dependency of these
differences: The amplitude of the difference between the RME of each
100\,\AA-bin and the RME of the reference region is plotted against
the central wavelength of each bin as full line. While this indicates
an signal of about 3\,m\,sec$^{-1}$ with a peak at the NaD doublet, it
should be noted that about half of the effect is due to the wavelength
dependency of the limb darkening. This can be seen from the dotted
line, where we keep the planetary radius constant at R$_{\rm
P}=1.31$\,R$_{\rm jup}$, the wavelength-dependent size corresponding
to the reference region. Here we can also see a peak at NaD, however,
with a lower amplitude (1.5\,m\,sec$^{-1}$). The observations of
\cite{2004MNRAS.353L...1S} indicate a measured amplitude of
1.7\,m\,sec$^{-1}$ for HD\,209458 from UVES spectra, in accordance to
our simulations.

As a possible strategy for an analysis we therefore suggest to use the
observed RME of the reference region to obtain the necessary stellar and
planetary parameters from a fit of the simulated RME (stellar radius,
orientation of the orbit with respect to the rotation axis of the star
as well as with respect to the line of sight and especially 
the planetary radius at this wavelengths) and use the corresponding
synthetic RME as reference. The differential effect as displayed in
the lower panels of Fig.\,\ref{Figrmnld2} and \ref{Figrmnld} would
then be the result of the analysis.

\subsection{The influence of the stellar rotation velocity}

\begin{figure}
\centering \includegraphics[width=8.8cm]{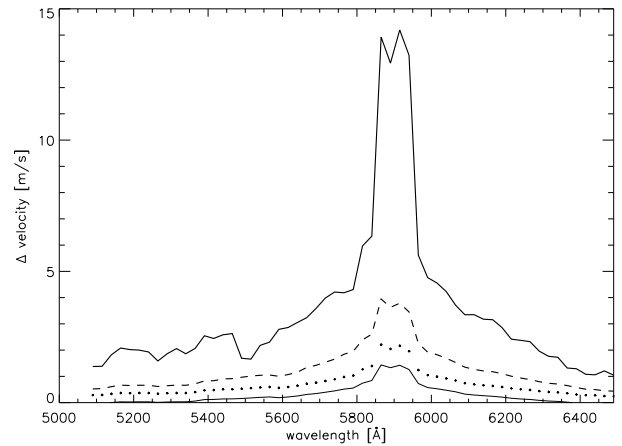}
\caption{The effect of stellar rotation: Same as lower panel of
  Fig.\,\ref{Figrmnld} but for stellar rotation velocities of v$_{\rm
  eq}=4.5, 6.0, 9.0, 25.0$\,km\,sec$^{-1}$ (thin, dotted, dashed, and
  thick line).}
\label{Figrmfast}
\end{figure}

For a broader application of these technique, the influence of the
rotation velocity of the host star is the most critical
parameter. While HD\,209458 is a slow rotator, several of the known
host stars of transiting planets rotate faster. In order to avoid
confusion between effects of changing stellar parameters for different
host stars, we compared the results from the previous paragraph now to
simulation for models where all parameters are kept fixed except the
rotation velocity. The results are presented in
Fig.\,\ref{Figrmfast}. As expected, the amplitude of the observable
effect increases with the rotation velocity of the host star. While
the amplitude is about 1.5\,m\,sec$^{-1}$ for a rotation velocity of
4.5\,km\,sec$^{-1}$ it increases to about 4\,m\,sec$^{-1}$ for a
rotation velocity of 9\,km\,sec$^{-1}$.  For planet host stars with
spectral types earlier than about F5 rotation is significantly larger
compared to those of later spectral type. Bright planet host stars of
earlier spectral types are therefore the primary targets for this
method.

\subsection{Wavelength-dependent transit light curves}

\begin{figure}
\centering \includegraphics[width=8.8cm]{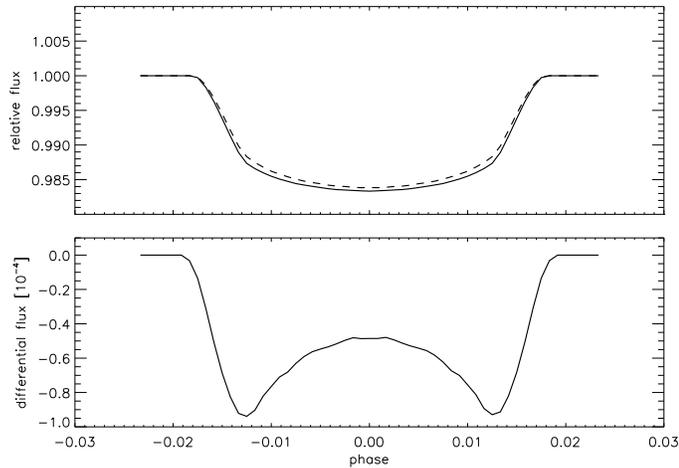}
\caption{Wavelength-dependent transit light curves: average from two
  wavelength regions centered around the two NaD doublet components of
  3\,\AA\, width each (top panel, full line), from a reference region
  5000-5200\,\AA\, (dashed line), and difference between the two
  (lower panel).}
\label{FigLight}
\end{figure}

As a by-product, we can also simulate transit light curves by
integrating the flux for each phase point. We can therefore also
provide detailed simulations for transmission photometry. As an
example, we show in Fig.\,\ref{FigLight} the transit light curve for
HD\,209458 for the reference region 5000-5200\,\AA\ (assuming a
box-like filter) and for the average of two 3\,\AA-bins each centered
around one of the NaD components. The difference between the two is
provided in the lower panel as possible observational result, similar
to those of \cite{2002ApJ...568..377C,2008ApJ...673L..87R,
2008A&A...487..357S, 2008ApJ...686..658S}. The differential effect is
largest during 
ingress and egress because the slightly larger planetary radius at the
NaD-lines causes a slightly early first and slightly later fourth
contact. The differential effect during phases between the second and
third contact is a combination of the slightly deeper transit due to a
larger radius at NaD and the differential effect of the limb darkening
between the two wavelength regions. This part of the curve therefore
depends critically on the wavelengths of the reference region.

Since (ground-based) measurements for the differential effect of the
transit light curves with wavelength regions of strong absorption
lines of the planetary atmosphere rely on high resolution and
high-quality spectra, both, the RME and the wavelength-dependent
transit 
light curve, can be evaluated simultaneously. With our simulations as
comparison, the detection efficiency can therefore be increased
significantly.

\section{Conclusion}

We performed simulations of the Rossiter-McLaughlin-effect using a
predicted wavelength dependency for the transiting planet in order to
evaluate the possibility to detect atmospheres of extrasolar planets
with ground-based high-resolution spectroscopy. We showed that a
detailed treatment of the spectrum synthesis of the partly eclipsed
stellar surface is important in order to detect these subtle
effects. In typical applications for the measurement of spin-orbit
alignment, we obtain systematic differences up to 6\,m\,s$^{-1}$ in
simulations tuned for HD\,209458. A faster rotation of the primary
star would increase these systematic effects.

The slightly larger planetary radius at the NaD-resonance line, caused
by additional absorption within the planetary atmosphere, produces a
slightly larger RME compared to a wavelength region where the
planetary atmosphere hast a very small contribution (e.g. the region
5000-5200\,\AA). Our simulations for HD\,209458, predict
an amplitude of the effect of 1.5\,m\,s$^{-1}$. This is a very small
effect which requires several phase-resolved high resolution spectra
from several transits taken with a stabilized spectrograph. While the
rotation velocity of HD\,209458 is low (4.5\,km\,s$^{-1}$), systems
with a primary of spectral type F5 or earlier typically show higher
rotation velocities. For rotation velocities of 25\,km\,s$^{-1}$ the
amplitude of the effect is already 15\,m\,s$^{-1}$. With increasing
rotational velocity, the accuracy of the radial velocity measurements
will, however, be reduced due to the broadening of the stellar lines.

From these simulation, we can also derive wavelength-dependent transit
light curves. A larger planetary radius at the NaD-resonance line
could be detected from a comparison between the light curve at this
wavelength to one at a reference region with small contributions
from the planetary atmosphere. 

%The application to phase-resolved,
%high-quality spectra would allow to probe the atmospheres of
%transiting planets with ground-based observations. 

During the recent years, atmospheres of extrasolar planets
  have been probed with different techniques. Photometric observations
  of the planetary albedo and secondary eclipses has to rely on
  space-based IR instruments, while the transit spectroscopy has
  recently been successful also from the ground (see Sect. 1). In this
  paper it has been been shown that the analysis of the
  wavelength-dependent Rossiter-McLaughlin-effect is an 
  alternative approach for a ground-based detection. While access from
  the ground will potentially enable a broader application, the
  information of the ground based transit spectroscopy as well as our
  proposed application of the RME is restricted to strong absorption
  lines in the planetary atmosphere. An advantage of the
  wavelength-dependent RME is that it can be obtained as a by-product
  of either ground-based transit spectroscopy, providing an
  independent detection of absorption lines in planetary atmospheres,
  or as by-product to RME measurements for spin-orbit alignment
  provided that the requirement for the wavelength stability of
  the spectrograph is high enough to measure the wavelength-dependent
  effect.

\begin{acknowledgements}
We thank P. Hauschildt for advice in computing the \texttt{PHOENIX}
intensity spectra. AR acknowledges financial support through an Emmy
Noether Fellowship from the Deutsche Forschungsgemeinschaft under DFG
RE 1664/4-1.
\end{acknowledgements}

\end{document}